\documentclass[12pt]{article}

\usepackage{graphicx}
\usepackage{epsfig}
\usepackage{psfrag}
\usepackage{wrapfig}
\usepackage[all]{xy}
\usepackage{bbm}

\usepackage{fullpage}
\usepackage{setspace}
\usepackage{url}
\usepackage{xcolor}
\usepackage{footmisc}
\newcommand\wordcount{
    \immediate\write18{texcount -sum -1 \jobname.tex > 'count.txt'}
\input{count.txt}words}

\usepackage{natbib}
\usepackage{amsmath}
\usepackage{amsfonts}
\usepackage{amsthm}
\usepackage{bbm}
\usepackage{array}
\usepackage{hyperref}

\def\E{{\rm E}\,}

\newcommand{\Y}{\mathbf{Y}}
\newcommand{\Z}{\mathbf{Z}}

\newcommand{\z}{\mathbf{z}}

\begin{document}

\author{Cyrus Samii, Ye Wang, Jonathan Sullivan, and  P.M. Aronow \thanks{ Samii (contact author) is Associate Professor, Department of Politics, New York University (Email: cds2083@nyu.edu). Wang is Assistant Professor, Department of Political Science, University of North Carolina-Chapel Hill (Email: yewang@unc.edu). Sullivan is Postdoctoral Fellow, School of Geography, Development, and Environment, University of Arizona (Email: jasullivan@arizona.edu).  Aronow is Associate Professor, Departments of Political Science and Biostatistics, Yale University (Email: p.aronow@yale.edu). }
}

\title{Inference in Spatial Experiments with Interference using the {\tt SpatialEffect} Package}

\maketitle
\pagenumbering{gobble}
\clearpage

\begin{center}
{\LARGE Inference in Spatial Experiments with Interference using the {\tt SpatialEffect} Package}
\end{center}
\vspace{5em}

\begin{abstract} 
This paper presents methods for analyzing spatial experiments when complex spillovers, displacement effects, and other types of ``interference'' are present.   
We present a robust, design-based approach to analyzing effects in such settings.
The design-based approach derives inferential properties for causal effect estimators from known features of the experimental design, in a manner analogous to inference in sample surveys.
The methods presented here target a quantity of interest called the ``average marginalized response,'' which is equal to the average effect of activating a treatment at an intervention node that is a given distance away, averaging ambient effects emanating from other intervention nodes.
We provide a step-by-step tutorial based on the {\tt SpatialEffect} package for R.  
We apply the methods to a randomized experiment on payments for community forest conservation in Uganda, showing how our methods reveal possibly substantial spatial spillovers that more conventional analyses cannot detect.  
\end{abstract}


\clearpage

\pagenumbering{arabic}
\doublespace
\section{Introduction}

\begin{figure}\centering 
\includegraphics[width=.75\textwidth]{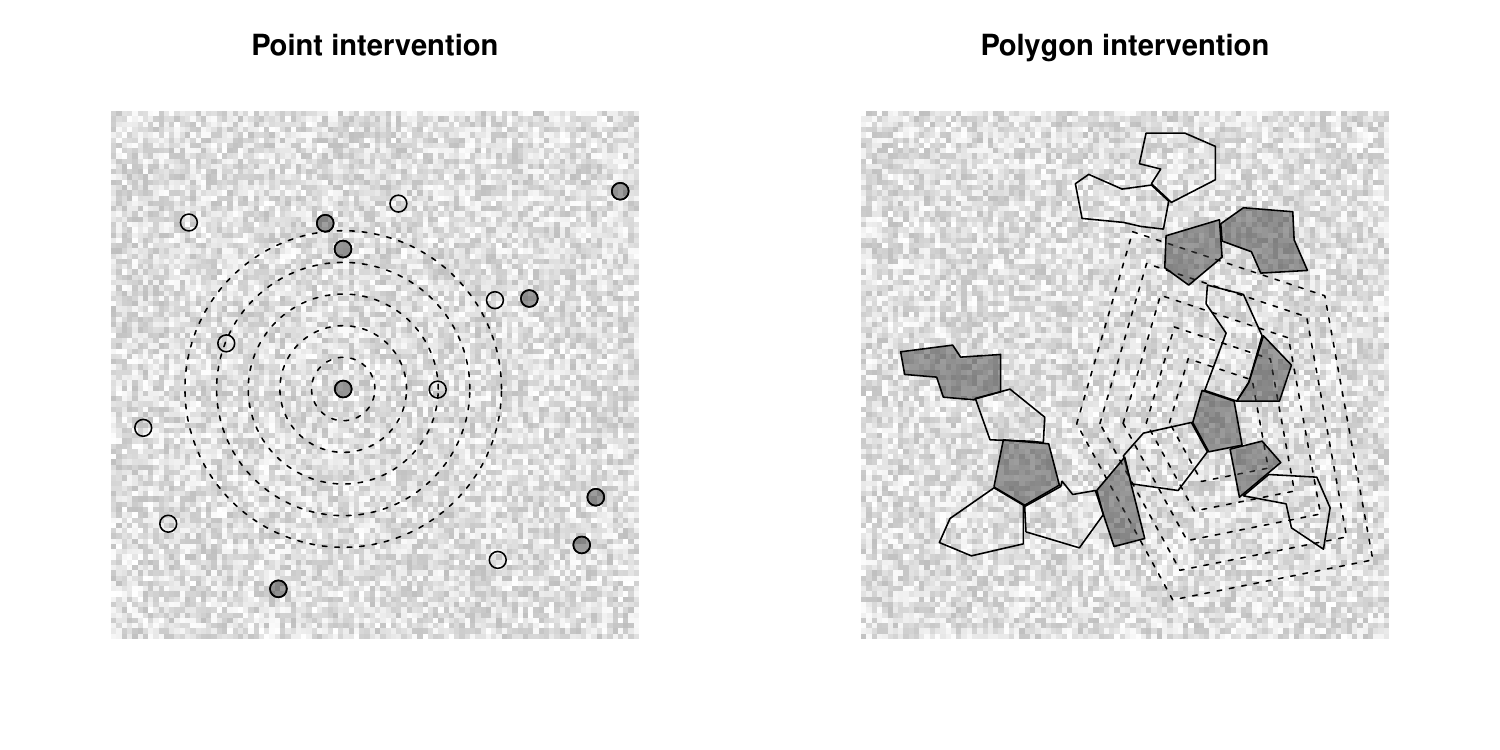}
\caption{Illustrations of hypothetical spatial experiments in which interventions are applied to points (left) or polygons (right). The background raster captures the geographic outcome data. We allow for the possibility for effects to bleed out in space, as illustrated by the concentric dashed lines. Figure reproduced from \citet{wang-etal-spatial}}
\label{fig:illustrations}
\end{figure}

Spatial experiments involve interventions that are applied to points or polygons in geographic space.
Such interventions have effects that transmit across space in potentially complicated ways. 
Figure~\ref{fig:illustrations} illustrates the generic structure of such experiments.
The left panel illustrates a potential point-intervention experiment. 
The points show locations at which an intervention might be applied.
An experimental design could assign half of such points to be treated with the intervention (gray shaded points), with the rest of the points remaining in the untreated, control condition (unshaded points). 
The shading in the background raster indicates outcome values.
The right panel illustrates a similar situation, but with treatments assigned to polygons instead of points.
A conventional analysis might work with outcomes within small circles surrounding the points or within the polygons, treating outcomes as independent, or at most, subject to some spatial correlation.
However, such modeling assumptions may be inadequate if the interventions have effects that bleed out in space.
The dashed lines illustrate possible zones into which effects may bleed out.
If effects bleed out in such ways, then outcomes by a given point or polygon depend not only on the treatment assignment of that point/polygon, but also on the pattern of treatment assignments for neighboring points/polygons.
Such spillover effects are examples of what statisticians, following \citet{cox58}, call causal ``interference.''  
This paper discusses an approach for analyzing spatial experiments when such interference is present.

Many environmental, health, agricultural, and other social interventions can be understood in these terms.
Consider, for example, ``payments for ecosystem services'' (PES) interventions, which typically offer money in exchange for recipients' refraining from degrading natural resources in their vicinity.  
Such an intervention may displace such degradation to more distant areas (negative spillover) or the intervention may have a bandwagon effect that induces people in neighboring areas to reduce degradation too (positive spillover) \citep{pfaff-2017-spillovers}. 
Ignoring negative spillovers can lead researchers to overstate the value of an intervention, while ignoring positive spillovers can lead researchers to understate this value.     
Below, we analyze spillovers using data from a PES experiment due to \citet{Jayachandran267}, finding indication of positive spillovers that the original analysis could not detect.

This paper presents an introduction to design-based inference for spatial effects when interference is present, drawing on the technical results presented in \citet{wang-etal-spatial}. Design-based inference is distinct from model-based inference \citep{sarndal_etal92}. Model-based inference is based on specifying a parametric or semi-parametric model for outcome variables, and then obtaining parameter estimates that best account for outcome variation. 
For spatial experiments, model-based approaches include spatial regression analysis \citep{arbia2006spatial, darmofal2015-spatial-book, kelejian_piras2017}.  Such methods are sensitive to misspecification. For example, a conventional modeling assumption is that effects are homogenous or that heterogeneity can be modeled in simple terms.  When this is not valid, researchers may misinterpret the generality of an estimated effect.

Design-based inference, by contrast, focuses on how well an experimental design or sampling design produces information about a target quantity under minimal assumptions on the outcome data generating process \citep{rubin2008-design-trumps}. A design based analysis is robust to a variety of outcome data generating processes and sources of effect heterogeneity.  
Design-based methods focus on carefully defined target quantities, such as average treatment effects, and the conditions for non-parametric estimation of such quantities. 
These quantities have immediate relevance for assessing the welfare implications of an intervention.
Under the stable unit treatment value assumption (SUTVA), the average treatment effect tells us the expected effect of assigning all units to treatment, regardless of the nature of effect heterogeneity \citep{imbens_rubin15}.
Unfortunately, interference directly violates SUTVA, and so this interpretation is no longer valid \citep{aronow2020spillover, savje-etal2018-unknown-interference}.
Interference motivates consideration of alternative quantities of interest.

Our analysis focuses on a quantity that we call the ``average marginalized response'' (AMR). 
The AMR measures how, on average, outcomes that at a given distance from an intervention site are affected by activating treatment at that site, taking into account ambient effects emanating from other intervention sites.  
From a policy perspective, the AMR tells us the average value of adding an additional treated site.
We can also use linear functionals of the AMR (e.g., the sum or the mean of AMRs) to characterize cumulative effects. 
In our illustrations below we demonstrate this.
Statistically, the AMR can be reliably estimated in spatial experiments through a simple contrast. 
\citet{wang-etal-spatial} show that when effects are additive or homogenous, the AMR is equivalent to the ``spillover" coefficient in a linear structural model.
The AMR retains a meaningful interpretation when effects are heterogenous or non-additive.  
We note that in targeting the AMR, we are not attempting to disentangle ``direct'' and ``indirect'' effects, as in analyses by \citet{hudgens_halloran08}, \citet{zigler-papadogeorgou2018-bipartite}, and \citet{hu2022average}.

The methods presented here are more agnostic about the structure of the interference than methods that specify an ``exposure mapping'' \citep{manski2012_identification_social, aronow_samii2017_interference}. We also work with looser assumptions on the spatial extent of interference as compared to methods based on ``partial interference.'' See \citet{aronow2020spillover} for a review of approaches based on exposure mappings and partial interference.

We apply these methods to an experimental study of a forest-conservation PES intervention in Uganda, originally analyzed by \citet{Jayachandran267}. 
We offer a step-by-step demonstration of the tools in the {\tt SpatialEffect} package for R \citep{samii_wang_spatial_package}, which was developed to implement the methods presented in \citet{wang-etal-spatial}.  
The required input data include raster or geographic point data on outcomes and then either geographic point data or polygon shapefile data on experimental treatments sites.

In the next section, we provide an accessible discussion of our methods, defining the AMR and strategies for inference.  Next, we describe how to implement this approach using the {\tt SpatialEffect} package for R, and use a small toy simulation to illustrate.   
We then turn to the forest conservation experiment in Uganda.

\section{Methods}

This section gives an overview of design-based inference for spatial effects. For a more detailed and technical treatment, we refer the reader to \citet{wang-etal-spatial}.

\subsection{Setting}

Our setting resembles the ``bipartite causal inference'' setup of \citet{zigler-papadogeorgou2018-bipartite}, in that we have a set of observation points and then a separately indexed set of intervention nodes.
As described at the outset, these intervention nodes could be points or polygons.
\citet{wang-etal-spatial} show that the analysis is essentially the same, and so for illustration here we will focus on point interventions.
Our application based on \citet{Jayachandran267} illustrates with a polygon intervention.
Both the observation points and intervention nodes reside in the same geographic space.  
Specifically, suppose that we have a set of intervention nodes indexed by $i=1,...,N$, and these nodes are arrayed as points in a spatial field, $\mathcal{X}$.  Each point in this spatial field, $x \in \mathcal{X}$, equals a coordinate vector of length 2 (recording, e.g., longitude and latitude coordinates).  
For point interventions, the coordinates of an intervention node $i$ can be given by $x(i)$.   

For each intervention node, we have a binary random variable $Z_i \in \{0,1\}$ that connotes whether, at node $i$, treatment is activated ($Z_i = 1$) or not ($Z_i = 0$). Denote the ordered vector of random treatment assignments as $\mathbf{Z} = (Z_1,...,Z_N)'$, with support $\mathcal{Z}$.  We define potential outcomes such that at each point $x \in \mathcal{X}$, we would have outcome $Y_x(\mathbf{z})$ when the treatment assignment is such that $\mathbf{Z} = \mathbf{z}$, and these potential outcomes are defined for each $\mathbf{z} \in \mathcal{Z}$. 
We stress that outcomes are defined at all $x$ values in $\mathcal{X}$, not just at the $x(i)$ values, which makes the setting bipartite.
This definition allows for the existence of interference: the potential outcome of each $x$ is decided not by the treatment status of any single intervention node (e.g., the nearest $Z_i$) but the treatment assignment vector for all the nodes (i.e., $\mathbf{Z}$).  The outcome that we observe at point $x$, labeled $Y_x$, is the potential outcome associated with the realized treatment assignment vector: 
$$
Y_x =  \sum_{\mathbf{z} \in \mathcal{Z}} Y_x(\mathbf{z}) \mathbf{1}(\mathbf{Z}=\mathbf{z}),
$$
where $\mathbf{1}(\cdot)$ is the indicator function.
Our approach does not require us to attempt to model how $Y_x$ is affected by the assignment vector $\mathbf{Z}$ in terms of a regression or other parametric specification (i.e., we allow for effects to be nonlinear and heterogenous).

The setup is consistent with how spatial experiments operate: $\mathcal{X}$ represents the area where the experiment occurs, for example a region, and the intervention nodes are places where the treatment could be potentially implemented, such as villages. $Y_x$ is the outcome---for example, whether a given point $x$ in the experimental area (e.g. region) is covered by forests or not. As discussed in the introduction, treatment effects in spatial experiments can have complex forms. Imposing structural restrictions such as homogenous spatial effects can lead to mischaracterizing the effect of an intervention.

Data for points in $\mathcal{X}$ come in various formats. We may have raster data for which the space is coarsened into raster cells.  Alternatively, we may have data for only a sparse set of points in $\mathcal{X}$, in which case we could either limit ourselves to working with those sparse points, or we could work with values interpolated between the points. 

Our analysis focuses on randomized experiments for which the distribution of the treatment assignment vector, $\Z$, is known.  Our inferential results are derived from the treatment assignment distribution.  Extensions to observational studies would follow if the treatment assignment mechanism resembled the kind of Bernoulli process assumed in \citet{wang-etal-spatial}.

\subsection{The average marginalized response}

For defining causal effects of interest, we rewrite the potential outcome at point $x$ as $Y_x(Z_i, \mathbf{Z}_{-i})$, where $\mathbf{Z}_{-i}$ is the treatment status of all the intervention nodes except for node $i$.  To concentrate on the effect generated by node $i$, we hold node $i$ to treatment value $z$ and marginalize over this variation in $\mathbf{Z}_{-i}$ to define an ``individualistic'' average of potential outcomes at point $x$:
$$
Y_{ix}(z; \alpha) = \textrm{E}_{\mathbf{Z}_{-i}}\left[Y_x(z, \mathbf{Z}_{-i}) \right] = \sum_{\mathbf{z}_{-i}\in \mathcal{Z}_{-i}} Y_x(z, \mathbf{z}_{-i})\mathrm{Pr}(\mathbf{Z}_{-i}=\mathbf{z}_{-i}; \alpha),
$$
where $\alpha$ is the experimental design parameter that governs the distribution of $\mathbf{Z}$ (that is, the probabilities of different treatment assignment vectors), $\mathbf{z}_{-i}$ is a vector of treatment values at nodes other than node $i$, and $\mathcal{Z}_{-i}$ is the set of possible values that $\mathbf{Z}_{-i}$ can take.  
The design parameter, $\alpha$, is an abbreviated way of capturing a variety of factors that might affect the distribution of $\mathbf{Z}$. 
For example, it could index vectors of treatment assignment probabilities for the intervention nodes, where such probabilities would determine the level of treatment saturation. 
It could also index the correlation structure for the treatments, which would characterize potential restrictions (for example, stratification or clustering) on the profiles of possible treatment assignments.

This allows us to define the causal effect at point $x$ of intervening on node $i$, allowing other nodes to vary as they otherwise would under $\alpha$:
$$
\tau_{ix}(\alpha) = Y_{ix}(1;\alpha) - Y_{ix}(0, \alpha).
$$
This defines the effect on raster cell $x$ of switching node $i$ from no treatment to active treatment, averaging over possible treatment assignments to the nodes other than $i$.

We can aggregate outcomes and effects at each intervention node as follows. First, define the ``circle average'' function:
$$
\mu_i(\Y(\z);d) = \frac{1}{|\{x:d_i(x)=d\}|}\int_{x:d_i(x)=d} Y_x(\z)\text{d}\zeta,
$$
where $d_i(x) = ||x(i) - x||$ is some well-defined metric that defines whether point $x$ is within a specified distance from intervention node $i$, which is located at $x(i)$, and $\zeta$ is a suitable integration measure.  
In the {\tt SpatialEffect} R package, the user has the possibility to define the distance metric. 
The default is Euclidean distance, in which case the circle average is defined in terms of the average of raster points that reside on the edge of a circle with radius $d$, defined in terms of the geographic coordinate values.
Another option that the {\tt SpatialEffect} package offers is to define the circle average as points that are within a $k$-coordinate-wide bandwidth around points that are $d$ coordinate units from the intervention node; this latter specification defines a disc or circular band (or ``donut'') of intervention nodes that are between $d-k$ and $d+k$ coordinate units from the intervention node. 
The quantity $\mu_i(\Y(\z);d)$ is the average of points selected with the specified distance metric.  
Then, 
$$
\mu_{i}(z; d, \alpha) = \E_{\Z_{-i}}[\mu_i(\Y(z, \mathbf{Z}_{-i}); d)]
$$
is the potential $d$-radius circle average around node $i$, given that node $i$ is assigned to treatment condition $z$ and marginalizing over possible assignments to other points.   The quantity $\mu_{i}(z; d, \alpha)$ is simply the circle average of the $Y_{ix}(z; \alpha)$ values. Finally,
$$
\tau_{i}(d; \alpha) = \mu_{i}(1; d, \alpha) - \mu_{i}(0; d, \alpha),
$$
is the average of individualistic responses for points along the circle of radius $d$ around node $i$.  
This is the circle average of the $\tau_{ix}(\alpha)$ values for outcome at distance $d$ from intervention node $i$.

Working with these terms, our target quantity is the {\it average marginalized response} (AMR) for distance $d$, which is simply the average $d$-radius effect over the intervention nodes:
$$
\tau(d; \alpha) = \frac{1}{N} \sum_{i=1}^N \tau_{i}(d; \alpha) 
.
$$
The interpretation of the AMR for distance $d$ is the average effect of switching a node $i \in \mathcal{N}$ to treatment on points at distance $d$ from that node, marginalized over possible realizations of treatment statuses in other intervention nodes.  The distribution of these possible realizations of treatment statuses depends on the experimental design. 

\subsection{Estimation and inference}

\citet{wang-etal-spatial} present non-parametric estimators for the AMR. On the basis of mean square error considerations, the preferred estimator for practical purposes is the Hajek estimator \citep[pp. 182-184]{sarndal_etal92}, given by:
\begin{align*}
\widehat{\tau}_{HA}(d) = \frac{1}{N_1}\sum_{i=1}^{N}Z_i \mu_i(\Y; d) - \frac{1}{N_0}\sum_{i=1}^{N}(1-Z_i) \mu_i(\Y; d) \label{eq:hajek}
\end{align*}
where $N_1 = \sum_{i=1}^{N}Z_i$ and $N_0 = N - N_1$. 

With raster data, we can estimate the $\mu_i(\mathbf{Y};d)$ values using outcome values from the raster cells. That is, with raster data, the value $Y_x$ corresponds to the value of the raster cell that lays over $x$. For data that consist of discrete points over a spatial field, we have two options. One is to use coarsened circle averages.  But if the discrete points are sparse, then another approach is to use interpolation.  The {\tt SpatialEffect} R package includes an approach to do this using a kriging fit.\footnote{The implementation is based on the {\tt Krig()} function in the {\tt fields} package for R.} Then, the predicted values from the kriging fit are used for the $Y_x$ values in the expression above. 
As discussed in \citet{wang-etal-spatial}, we can also smooth the AMR estimates using a local regression approach.

\citet{wang-etal-spatial} show that under a set of mild regularity conditions, the Hajek estimator is consistent and asymptotically normal. Essentially, the conditions require the intervention nodes are separated from each other and there exists an upper bound on the number of nodes that interfere with each other. 
This allows for two types of inference for the estimated AMR. 
The first is a ``sharp null hypothesis'' permutation test.  
The sharp null hypothesis states that the interventions have no effects whatsover, such that $Y_x(\Z) = Y_x(\mathbf{0})$ , in which case all unit-level effects are also zero and the AMR is zero.
If one assumes that the sharp null hypothesis holds, then the randomization distribution of the AMR can be simulated by simply permuting treatment assignment, estimating the AMR under each of these permutations, and then collecting the resulting set of estimates \citep[Ch. 5]{imbens_rubin15}.
The resulting distribution is referred to as the null permutation distribution.
For a two-way test of the sharp null hypothesis for a specific AMR value with an error rate of $\delta$, one can evaluate the estimated AMR relative to the $\delta/2$ and $1-\delta/2$ quantiles of the null permutation distribution. 
(We use $\delta$ for the inferential error rate here so as not to confuse with $\alpha$, which is defined above to characterize the experimental design.)
Such tests of the sharp null are approximately in exact finite samples, where the quality of the approximation depends the extent to the which the simulated permutations cover the full set of possible permutations.
We can also use a permutation test to test linear functionals of the AMR, such as the mean or cumulative AMR within a range of distance values.
To do so, one evaluates the estimated AMR functional value against the null permutation distribution of the functional values.
We illustrate this approach below.  

The second type of inference provides confidence intervals that combine a normal approximation with a non-parametric spatial heteroskedasticity and autocorrelation consistent (spatial-HAC) standard error estimator as proposed by \citet{conley99_spatial}. 
These intervals characterize uncertainty for the estimated AMR, and can be used to test the ``weak null hypothesis'' that the AMR equals zero. 
(The weak null is implied by the sharp null, but the reverse does not hold.)
The {\tt SpatialEffect} package allows one to construct confidence intervals in either a pointwise manner, for specific AMR values, or in the form of a confidence band for simultaneous inference on the entire AMR curve. 
\citet{wang-etal-spatial} describe conditions under which these intervals are asymptotically conservative with respect to type-I error.  

\section{Feature of the {\tt SpatialEffect} package}

We have developed an R package, {\tt SpatialEffect}, to implement the methods described above. 
As we focus on bipartite designs, the key command in  package, also called {\tt SpatialEffect()}, takes two main inputs. 
The first is the data of the intervention nodes that records both their coordinates and treatment status. It takes the form of either a 
dataset with point intervention coordinates and attributes that include treatment status ({\tt Zdata}) or, for polygon interventions, a shapefile or a raster object ({\tt ras\_Z}) with polygons giving the location of each node along with attributes that include treatment status.
The second is the outcome data, which can also be either a raster object ({\tt ras}) or a dataset with the location and outcome value of each outcome point ({\tt Ydata}). Users need to specify the variable names of location ({\tt x\_coord\_Z}, {\tt y\_coord\_Z}) and treatment ({\tt treatment}) in {\tt Zdata} as well as location ({\tt x\_coord\_Y}, {\tt y\_coord\_Y}) and outcome ({\tt outcome}) in the outcome data if no raster object is provided.\footnote{If neither {\tt ras} nor {\tt Ydata} is provided, the command will try to find {\tt outcome} in {\tt Zdata} and interpolate the outcome values on $\mathcal{X}$ using kriging.}

Users also need to specify the vector of distance values ({\tt dVec}) at which the AMR will be estimated. It is usually helpful if users plot the data first to check the unit and range of distance in the sample. 
The distance metric ({\tt dist.metric}) can be based on Euclidean distance or geodesic distance. Users can choose a set of points that form the edge of a circle of radius $d$, that form a circular band (``donut'') centered at radius $d$, or that form a sector centered at $d$ (``disk''), by altering the option {\tt cType}.
When {\tt ras\_Z} is provided, the distance will be calculated from the boundary of each polygon and we implement polygon intervention. Otherwise, it is calculated from the center of each intervention node and point intervention is implemented.
There are also several parameters that allow users to control the precision when constructing the circle averages. {\tt numpts} is the number of sampling points on a circle. If no value is submitted, the number will be decided by the raster object's resolution. We sample roughly the same number of points for circles with different radius.  Users can also choose whether to sample each outcome point only once by setting the value of {\tt only.unique}.


As discussed in the previous section, the package provides users with two inference methods, a Fisher-style permutation test ({\tt per.se}) and an asymptotic variance estimator based on Conley (1999) ({\tt conley.se}). 
Users can change the error rate $\delta$ defining the level of statistical significance ({\tt conley.se}). 
The permutation test can be adjusted to account for blocking or clustering in experimental design. For the asymptotic variance estimator, users need to specify a distance threshold ({\tt cutoff}) beyond which interference dependence does not exist.  This can be based on substantive judgments about the likely spatial extent of interference, and can be varied as part of a sensitivity analysis. Users have the discretion on which kernel ({\tt kernel}) to use when calculating the standard errors\footnote{There are three kernels available: the uniform kernel (``{\tt uni}" or ``{\tt uniform}"), the triangular kernel (``{\tt tri}" or ``{\tt triangular}") and the Epanechnikov kernel (``{\tt epa}" or ``{\tt epanechnikov}").} and whether to apply an effective degrees of freedom adjustment ({\tt edf}) as proposed by \citet{young-2015-improved-inference}\footnote{The adjustment rescales the variance estimate such that its distribution is a more accurate approximation to the chi-square distribution.}. If the option {\tt smooth} is set to 1, the output will also contain the smoothed AMR estimates.

The command returns a list of objects. The first element in the list, {\tt AMR\_est}, is a data frame that contains each value of {\tt dVec} and the corresponding estimate of the AMR at that distance value. If {\tt per.se} is set to be 1, then the list includes an element named {\tt Per.CI}. 
For a specified error rate of $\delta$ (e.g., $\delta = 0.05$), it is the $\delta/2$ and $1-\delta/2$ percentiles of the sharp null distribution generated by permutation. 
At each distance $d$, the sharp null hypothesis of no effect is rejected if the AMR estimate is smaller than the $\delta/2$ percentile or larger than the $1-\delta/2$ percentile. {\tt Conley.SE} and {\tt Conley.CI} are in the list if {\tt conley.se} equals 1. 
For an error rate $\delta$ , they are the spatial HAC standard errors and the corresponding $1-\delta$ confidence intervals for the AMR estimate at each distance $d$, respectively. The smoothed AMR estimates (called {\tt AMR\_est\_smoothed}) is in the list if the option {\tt smooth} is turned on. Depending on the values of aforementioned options, the list may also include measures of their uncertainties ({\tt smoothed.Conley.SE} and {\tt smoothed.Conley.CI}).

Another command in the package, {\tt SpatialEffectTest()}, helps researchers evaluate the significance of linear functionals of the estimated AMR effects via a Fisher-style permutation test. 
It takes the output from {\tt SpatialEffect()} as an input and needs users to decide the distance range on which the effects are aggregated in the form of a linear functional (i.e., scaled sum) of the AMR estimates.
This can be used to estimate a mean effect or a cumulative effect, in the form of a Riemann sum approximation of the area under the AMR curve. 
The {\tt SpatialEffectTest()} function generates the null permutation distribution of the sum of the AMR estimates within the specified range. 
One can then use this null distribution to test the statistical significance of the linear functional estimate.

We also have display functions {\tt summary()} and {\tt print()}. 
Both take the output of {\tt SpatialEffect()} as the input and print estimates within the distance range specified by the user ({\tt dVec.range}).

\section{A toy example}
\begin{figure}\centering 
\includegraphics[width=0.4\textwidth, height=0.4\textwidth]{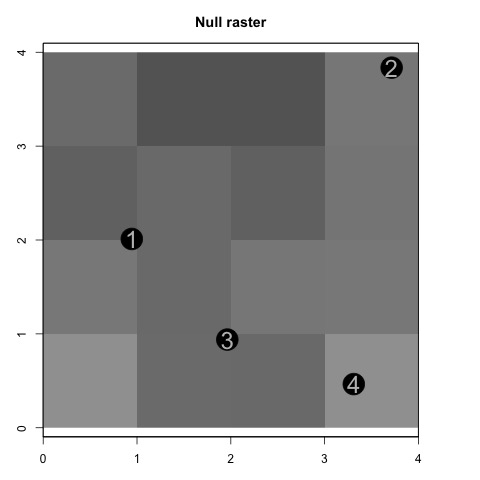}
\includegraphics[width=0.4\textwidth, height=0.4\textwidth]{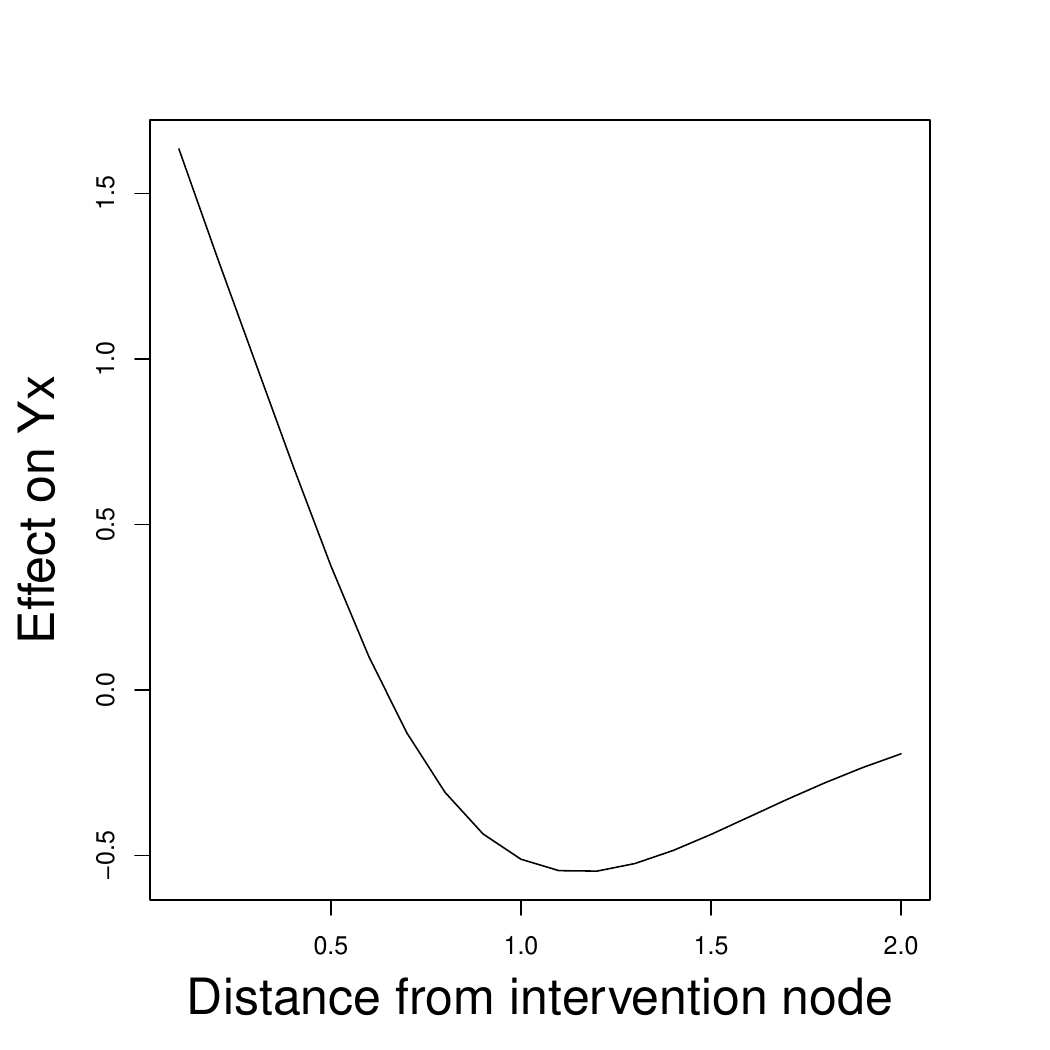}
\caption{The plot on the left displays the 4-by-4 raster that records the outcome when treatment is absent ($Y_x(0)$). Black points are the intervention nodes. The plot on the right presents the non-monotonic effect function.}
\label{fig:effect}
\end{figure}

We now illustrate concepts discussed above using a toy simulated example based on point interventions. 
We offer a full-fledged application with polygon interventions in the next section.
The toy example includes 4 intervention nodes in a spatial field where outcomes are recorded in a 4-by-4 raster (left plot of Figure~\ref{fig:effect}). We construct a hypothetical effect function by mixing two gamma-distribution kernels. This yields a non-monotonic effect emanating from each intervention node, which is presented on the right plot of Figure~\ref{fig:effect}.

To get the true AMR, we need to marginalize over all of the ways that treatment could be applied. We suppose Bernoulli random assignment of the nodes to treatment, in which case there are 16 ways treatment could be applied, all with equal probability. We then simulate the potential outcomes that would correspond to each treatment assignment and view the result.

\begin{figure}\centering 
\includegraphics[width=.8\textwidth]{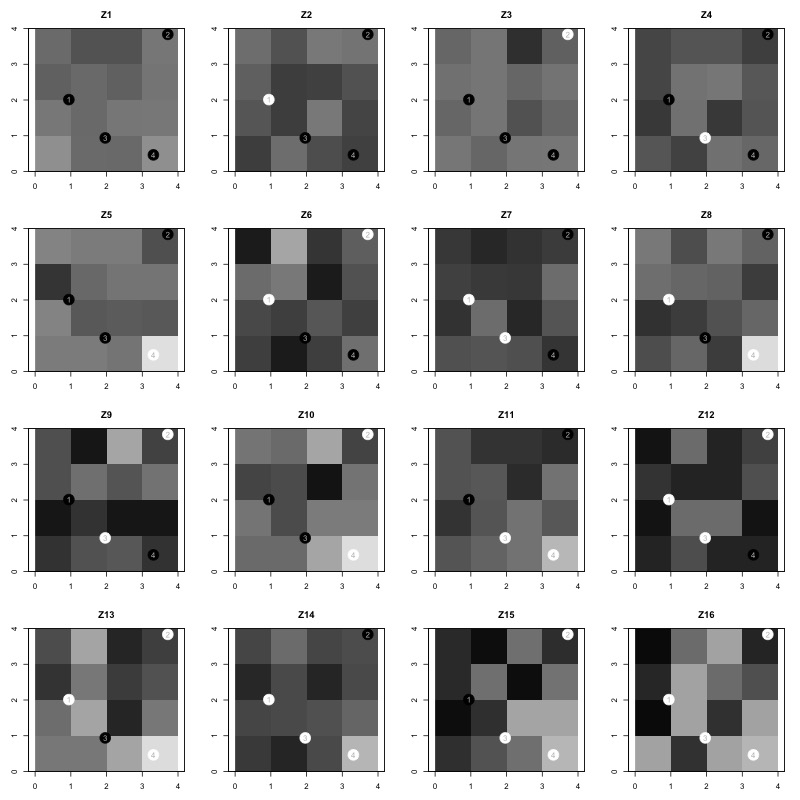}
\caption{The plots show the outcome's values in the raster under each of the sixteen possible assignments. Black nodes are treated and white ones are under control.}
\label{fig:raster}
\end{figure}

We can construct the causal quantities using outcomes under the sixteen assignments. First are the values of $\tau_{ix}(\alpha)$ (the node-and-point-specific effects). For each grid cell, there will be a separate $\tau_{ix}(\alpha)$ value corresponding to the effect of switching treatment status of each intervention node, averaging over possible assignments at other nodes. This corresponds to the difference in mean potential outcomes when node $i$ is in treatment versus in control, where these means are taken with respect to the set of assignments in the assignment matrix ($\mathcal{Z}$ in the formal analysis). So, to compute the values, we go through each assignment and take differences in means for outcome values corresponding to treatment status with 1 minus those with 0. For example, to calculate $\tau_{1x}(\alpha)$, we take the average over outcome values in graphs where node 1 is treated and subtract from it the the average over outcome values in graphs where node 1 is under control.

\begin{figure}\centering 
\includegraphics[width=0.4\textwidth, height=0.4\textwidth]{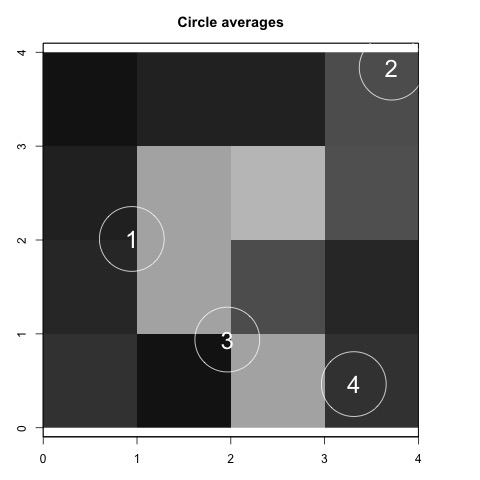}
\includegraphics[width=0.4\textwidth, height=0.4\textwidth]{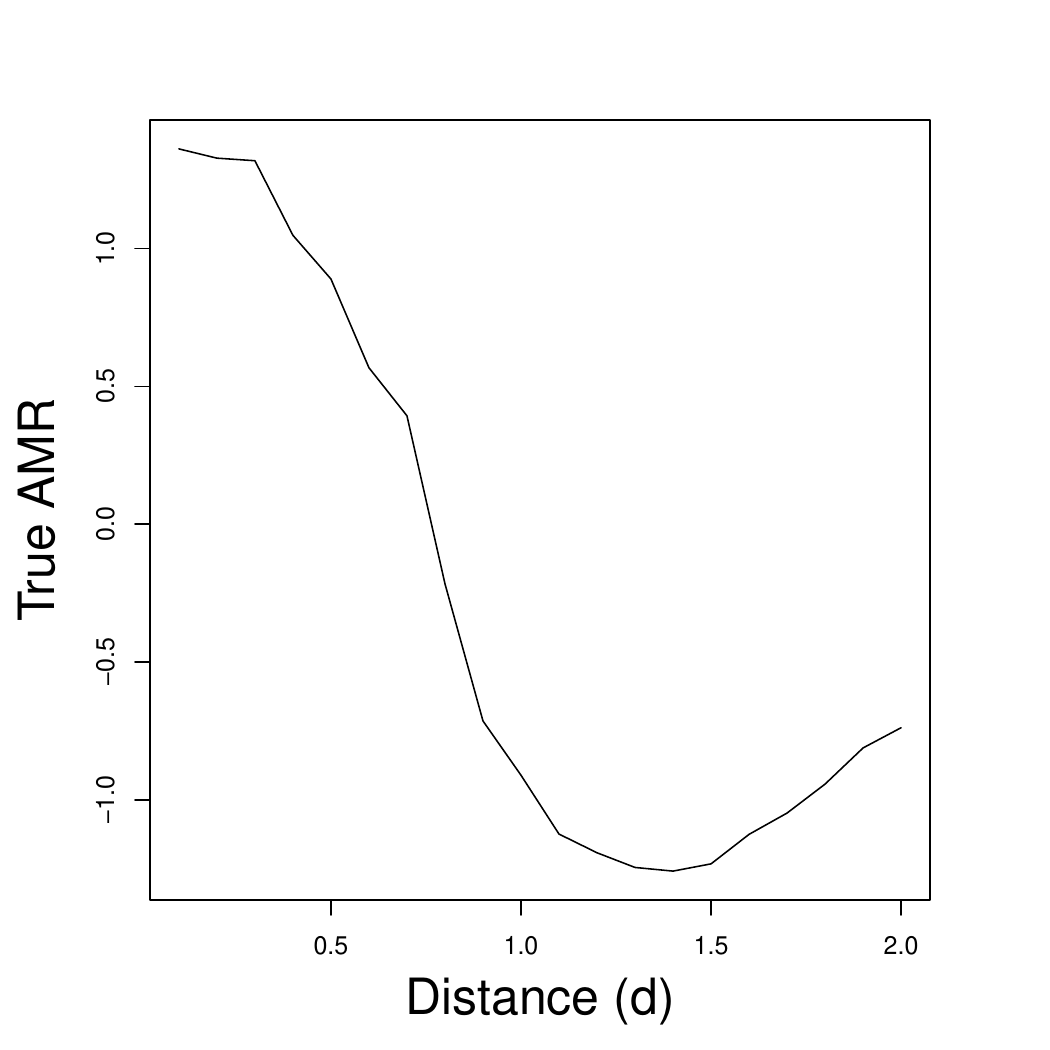}
\caption{The plot on the left displays the construction of circle averages around each intervention node. The circles are drawn with radii equal to a distance of $d=0.35$.  The plot on the right shows the true AMR curve in the toy example.}
\label{fig:AMR}
\end{figure}

We then obtain the $d$-specific $\tau_{ix}(\alpha)$ values for each of the intervention nodes by drawing circles with radius $d$ around them. The left plot of Figure~\ref{fig:AMR} shows how the circle averages are constructed when the distance metric is Euclidean distance and $d=0.35$. For this particular $d$, the circle average for node $1$ uses four grid cells while that of node $2$ only uses one. Finally, we aggregate these values to get the vector of $d$-specific AMR. The true AMR values reflect the non-monotonicity of the effect function (the right plot of Figure~\ref{fig:AMR}).

\section{Spatial effects in the toy example}
The first step in the analysis is to prepare {\tt Zdata}. We extract the location and treatment status of our four intervention nodes under a particular assignment. We name the variables {\tt x}, {\tt y}, and {\tt Z}. We also name the raster object in the toy example {\tt ras\_sim}. Based on the plot, we choose {\tt dVec} to be a sequence from $0.1$ to $2$ with an interval length of 0.1. The cutoff point for the spatial HAC variance is set to be 0.4 and the number of permutations is set to be 1000.\footnote{The number doesn't make any difference here as there are in total 6 assignments.}

We use the {\tt SpatialEffect()} function to estimate the AMR as follows: 
\begin{center}
\begin{verbatim}
dVec = seq(from=.1,to=2, by=.1)
AMRest <- SpatialEffect(ras = ras_sim, Zdata = Zdata, x_coord_Z = "x",  
                        y_coord_Z = "y",  treatment = "Z",  dVec = dVec, 
                        cutoff = .4, nPerms = 1000, smooth = 1)
\end{verbatim}
\end{center}

Next, we summarize the estimates when the distance values are between 0.1 and 1 with either {\tt summary()} and {\tt print()}. {\tt Conley.CI.l} and  {\tt Conley.CI.u} refer to the lower and upper bounds of the confidence intervals based on the spatial HAC variance estimator. {\tt Per.CI.l} and  {\tt Per.CI.u} are bounds of the confidence intervals under the null distribution.
\begin{center}
\begin{verbatim}
summary(AMRest, dVec.range = c(0.1, 1))
\end{verbatim}
\end{center}

\begin{center}
\begin{verbatim}
      dVec AMR_est Conley.CI.l Conley.CI.u Per.CI.l Per.CI.u
 [1,]  0.1   1.491      -0.277       3.259   -1.566    1.566
 [2,]  0.2   1.550      -0.242       3.343   -1.626    1.626
 [3,]  0.3   1.550      -0.242       3.343   -1.626    1.626
 [4,]  0.4   1.691      -0.050       3.431   -1.666    1.666
 [5,]  0.5   1.714      -0.020       3.448   -1.684    1.684
 [6,]  0.6   1.700      -0.038       3.438   -1.673    1.673
 [7,]  0.7   1.678      -0.066       3.422   -1.657    1.657
 [8,]  0.8   1.002      -0.067       2.071   -1.005    1.005
 [9,]  0.9  -0.816      -2.424       0.791   -1.328    1.328
[10,]  1.0  -0.791      -2.295       0.714   -1.307    1.307
\end{verbatim}
\end{center}

\begin{figure}\centering 
\includegraphics[width=0.5\textwidth]{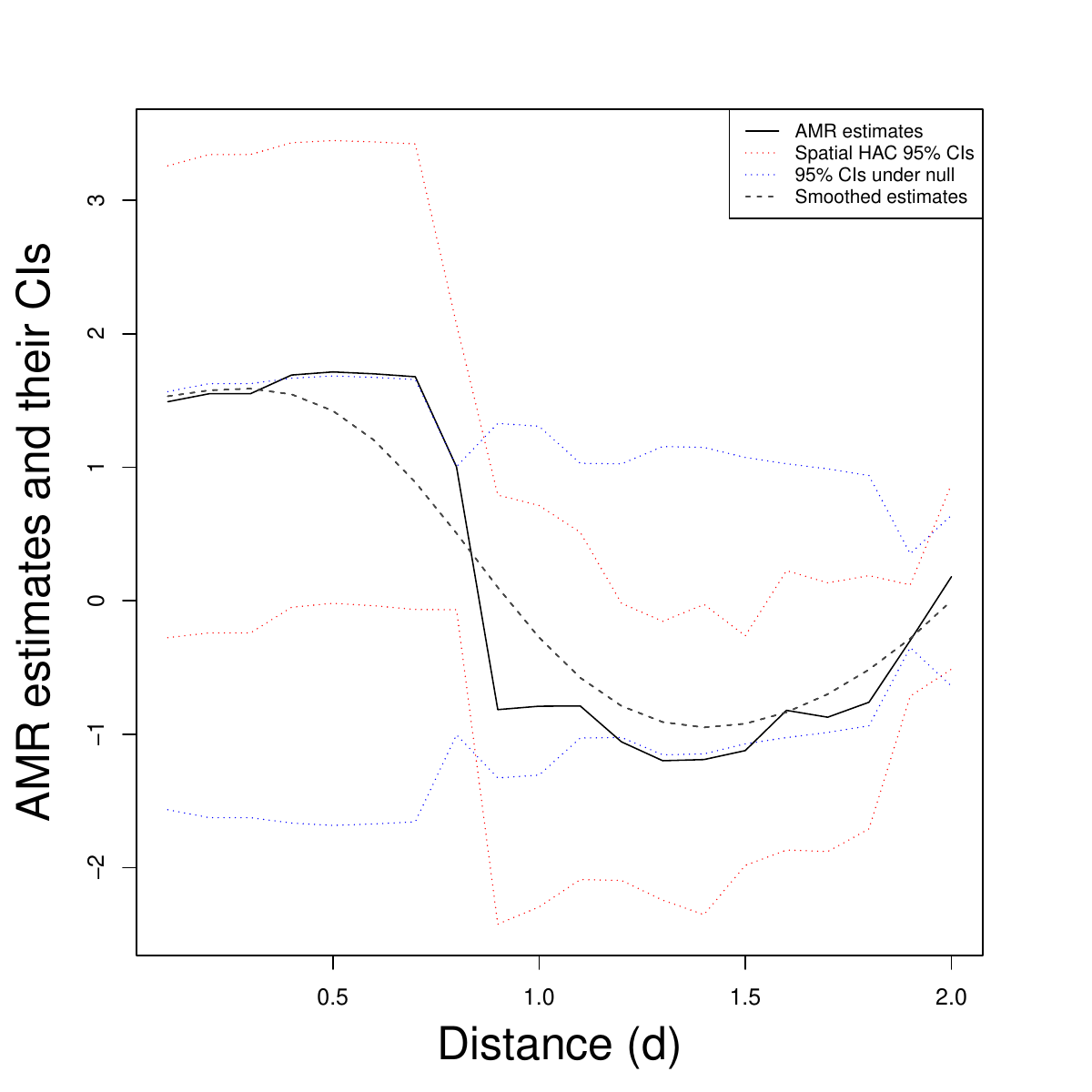}
\caption{The black solid line in the graph is the AMR estimate for each of the distance values $d$. The red dotted line is the 95\% confidence interval based on the spatial-HAC standard errors. The blue dotted line shows the 0.025 and .975 percentiles of the sharp null permutation distribution. The black dashed line is the smoothed AMR estimates. }
\label{fig:toy-results}
\end{figure}

The results are also presented in Figure~\ref{fig:toy-results}. The black solid line in the graph is the AMR estimate for each of the distance values $d$. The red dotted line is the 95\% confidence interval based on the spatial-HAC standard errors. The blue dotted line shows the 0.025 and .975 percentiles of the sharp null permutation distribution. The black dashed line is the smoothed AMR estimates. The results suggest that the AMR estimates are marginally significant when $d$ is small, where the estimates are slightly larger than the .975 percentile of the sharp null permutation distribution and the red line on the bottom is close to be positive.
We also see that the estimated AMR corresponds to the shape of the AMR in Figure~\ref{fig:AMR}.

To further test this initial inference regarding effects at small values of $d$, we use {\tt SpatialEffectTest()} to examine whether the cumulative effect is statistically significant on the interval $[0.1, 0.5]$:
\begin{center}
\begin{verbatim}
testResult <- SpatialEffectTest(AMRest, c(0.1, 0.5))
\end{verbatim}
\end{center}
The 0.025 and .975 percentiles of the sharp null permutation distribution of the cumulative effect are $-7.97$ and $7.97$, respectively, while the actual cumulative effect is $12.38$.  Based on an error rate of $\delta = 0.05$, we would reject the sharp null hypothesis that there is no cumulative effect on the distance interval of $[0.1, 0.5]$.

\citet{wang-etal-spatial} present simulation studies to show how finite sample inference converges toward expected asymptotic performance in the number of intervention nodes.
Computation time depends on whether one is smoothing the AMR curve, and if so whether one uses cross-validation to calibrate the smoother bandwidth, whether one is working with point interventions or polygon interventions, and then the resolution of the spatial polygons or rasters.
For example, on a MacBook with a 2.4 GHz dual-core processor, with point interventions, an outcome raster resolution of 100 cells per intervention node, and sample sizes of 16, 64, and 144 intervention nodes, run time is approximately 12, 54, and 78 seconds without cross-validation-calibrated smoothing, and 17, 79, and 238 seconds with cross-validation-calibrated smoothing.
For the forest conservation application that we present below, which uses polygon interventions, high resolution spatial data, and cross-validation-calibrated smoothing, the run time is approximately 2.5 hours.

\section{An experiment on payments for forest conservation}

We demonstrate the utility of the {\tt SpatialEffect} package by evaluating a payments for ecosystems services (PES) conservation program in Uganda. 
PES programs are a form of environmental governance that financially renumerate landowners for environmental performance, such as reduced forest loss, and have become a cornerstone of international climate policies aimed at reducing emission from forest loss \citep{agrawal-2011-reducing, ferraro-2011-future, ferraro-kiss-2002}. 
Prior assessments of PES programs used quasi-experimental methods to estimate the effects but have found divergent results based on selected designs and control
strategies \citep{alix-garcia-2015-only, Arriagada-2012-payments, Pattanayak-2010-show, samii2014effects}.
While evaluation methods of conservation programs have increased in recent years, the knowledge of potential spillover effects beyond the immediate location of intervention sites remains scarce \citep{pfaff-2017-spillovers}.

\begin{figure}\centering 
\includegraphics[width=0.4\textwidth]{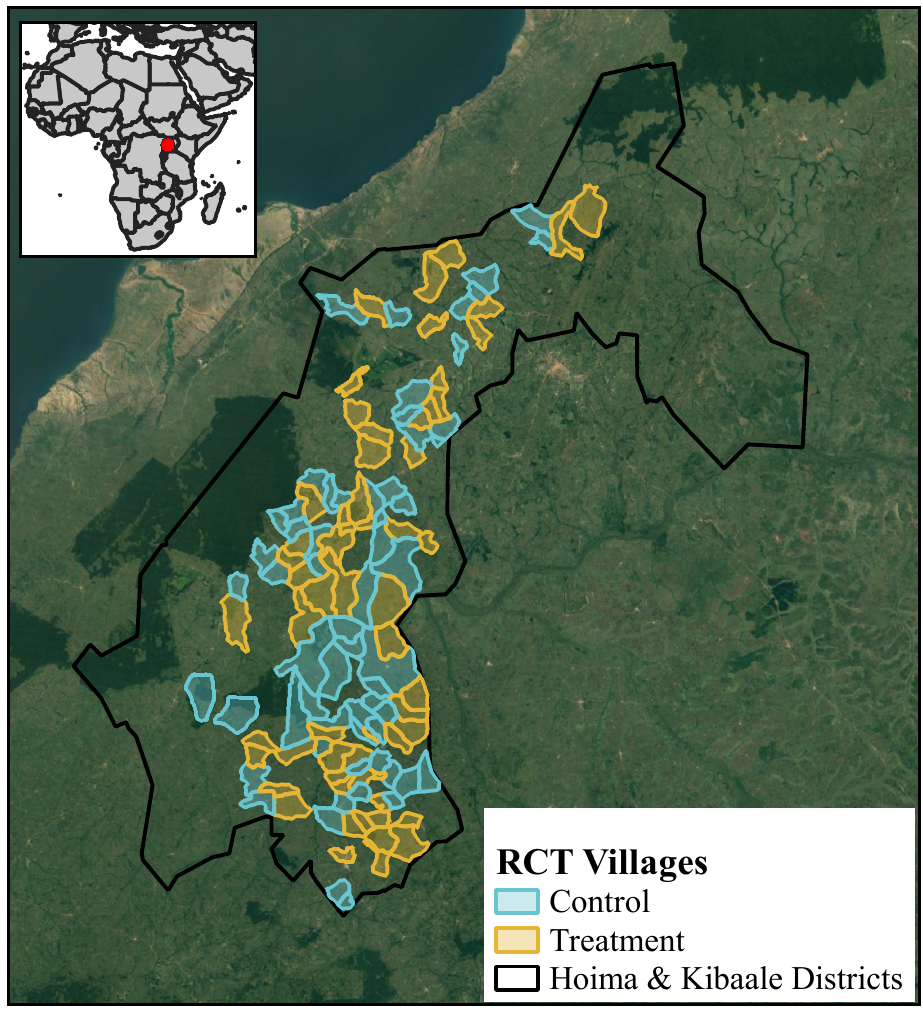}
\caption{Study area of randomized control trial for a PES program in Hoima and Kibaale district in Uganda, from \citet{Jayachandran267}. Boundaries of treatment (60) and control (61) villages were digitized using publicly available data and published maps.}
\label{fig:jaya-map}
\end{figure}

The PES case study that we examine in Uganda was previously evaluated by \citet{Jayachandran267} with a randomized control trial design, a rarity in conservation policy evaluation \citep{wiik-2019}.
The PES program was designed and implemented by a local nonprofit---Chimpanzee Sanctuary and Wildlife Conservation Trust (CSWCT)---in Hoima and northern Kibaale districts of Uganda (Figure \ref{fig:jaya-map}). 
The study was carried out in 121 villages, 60 of which were randomly assigned to treatment (Figure \ref{fig:jaya-map}). In treatment villages, the PES program was advertised to private forest owners (PFOs) to enroll in a 2-year conservation program from 2011 to 2013. 
Enrollees could receive 70,000 Ugandan shillings (UGX), or \$28 in 2012 US dollars, per hectare of forest per year. 
The PES program achieved a low-level of enrollment with 180 private forest owners (32\%) participating but a high-level of compliance with 88\% of enrollees meeting the requirements to preserve forested land \citet{Jayachandran267}. Individual behavioral changes were found to aggregate to village level landscape differences. 
In treatment villages, tree cover, estimated using high-resolution satellite imagery, declined 4.2\% over the study period compared to 9.1\% in control villages. 

A common concern of PES programs is whether conservation in one region may accelerate deforestation elsewhere due to equilibrium effects or non-compliant participants who shift to deforesting beyond regulated forest boundaries---so-called ``leakage.'' \citet{Jayachandran267}'s main analyses take the villages in the study to be independent, assuming no spillover. They address the possibility of spillover only to a limited extent.  First, they perform an indirect test by studying whether the effects of PES were larger in areas adjacent to forest reserves, given that such locations are places where displacement might be more likely. They do not find such effect heterogeneity to be present.  Second, they look at whether forest loss in control villages varied in terms of the number of treated villages within 5km, finding no evidence for such a pattern.  As the authors recognize, these two tests still allow for the possibility for spillover effects in the broader forest system outside the treated and control villages themselves, a possibility that is visibly apparent in looking at how the villages are situated in Figure~\ref{fig:jaya-map}.  Moreover, the result with regard to the number of nearby treated villages could be sensitive to the choice of 5km bandwidth. (For a related discussion, see \citet{clemens2015mapping}.)

\begin{figure}\centering 
\includegraphics[width=.8\textwidth]{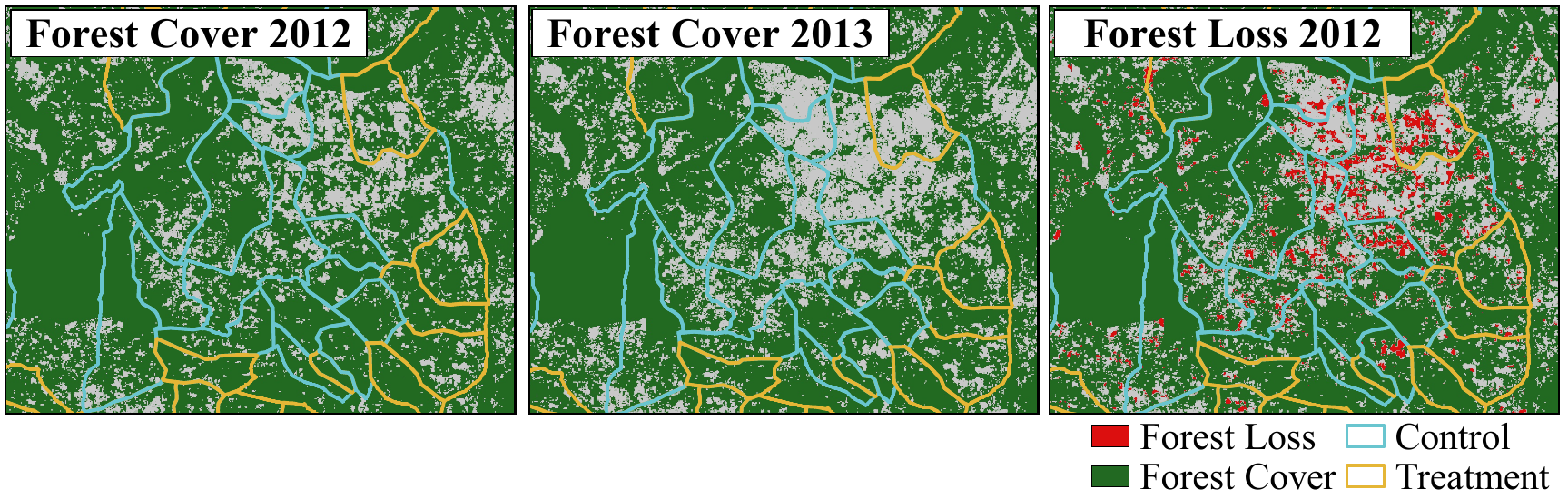}
\caption{The Global Forest Cover (GFC) dataset over a subset of the study area showing forest cover for 2012, 2013, and forest loss in 2012 \citep{Hansen850}.}
\label{fig:jaya-forests}
\end{figure}

To characterize spillovers in a more robust manner, we extend the analysis to include forest raster data from the entire region (not just within the treated and control villages) and apply the methods described above to estimate the AMR curve. 
The data used for our reanalysis includes the treatment and control village boundaries from the original \citet{Jayachandran267} study. 
The boundaries were not publicly available and thus hand digitized from available spatial data and published maps.\footnote{The original data repository can be found at: \\https://dataverse.harvard.edu/dataset.xhtml?persistentId=doi:10.7910/DVN/MGMDYN} 
The data is stored as a shapefile named {\tt ras\_Z}.
The forest cover data available from \citet{Jayachandran267} was limited to village boundaries. 
Therefore we use the \citet{Hansen850} forest cover dataset for years 2012 and 2013 with greater than 25\% tree cover over the study area. 
Figure~\ref{fig:jaya-map} shows our reconstructed map of the study area, and Figure~\ref{fig:jaya-forests} zooms in to a subset of the study area to show forest cover values from \citet{Hansen850} that we used in the analysis. 
The data is stored as a raster object named {\tt ras}.

We set the range of distance values to be between 0 kilometer and 20 kilometers. We consider the experiment as an example of polygon interventions. Therefore, the starting point at 0 corresponds to the area within the boundary of each village. It is worth noting that the distance unit in the raster of forests is decimal degrees. Since 1 decimal degree is approximately 111 km, the distance values we select are between 0 and $20/111 \approx 0.18$, with a step size of $5/111 \approx 0.045$. The cutoff is set at 10 km = 10,000 m. We specify the {\tt SpatialEffect()} function as follows:

\begin{center}
\begin{verbatim}
dVec = seq(from=0, to=20/111, by=5/111)
AMRest <- SpatialEffect(ras = ras, ras_Z = ras_Z,  treatment = "treatment",  
                        dVec = dVec, smooth = 1, cutoff = 10000, 
                        dist.metric = "Geodesic")
\end{verbatim}
\end{center}

\begin{figure}\centering 
\includegraphics[width=0.4\textwidth]{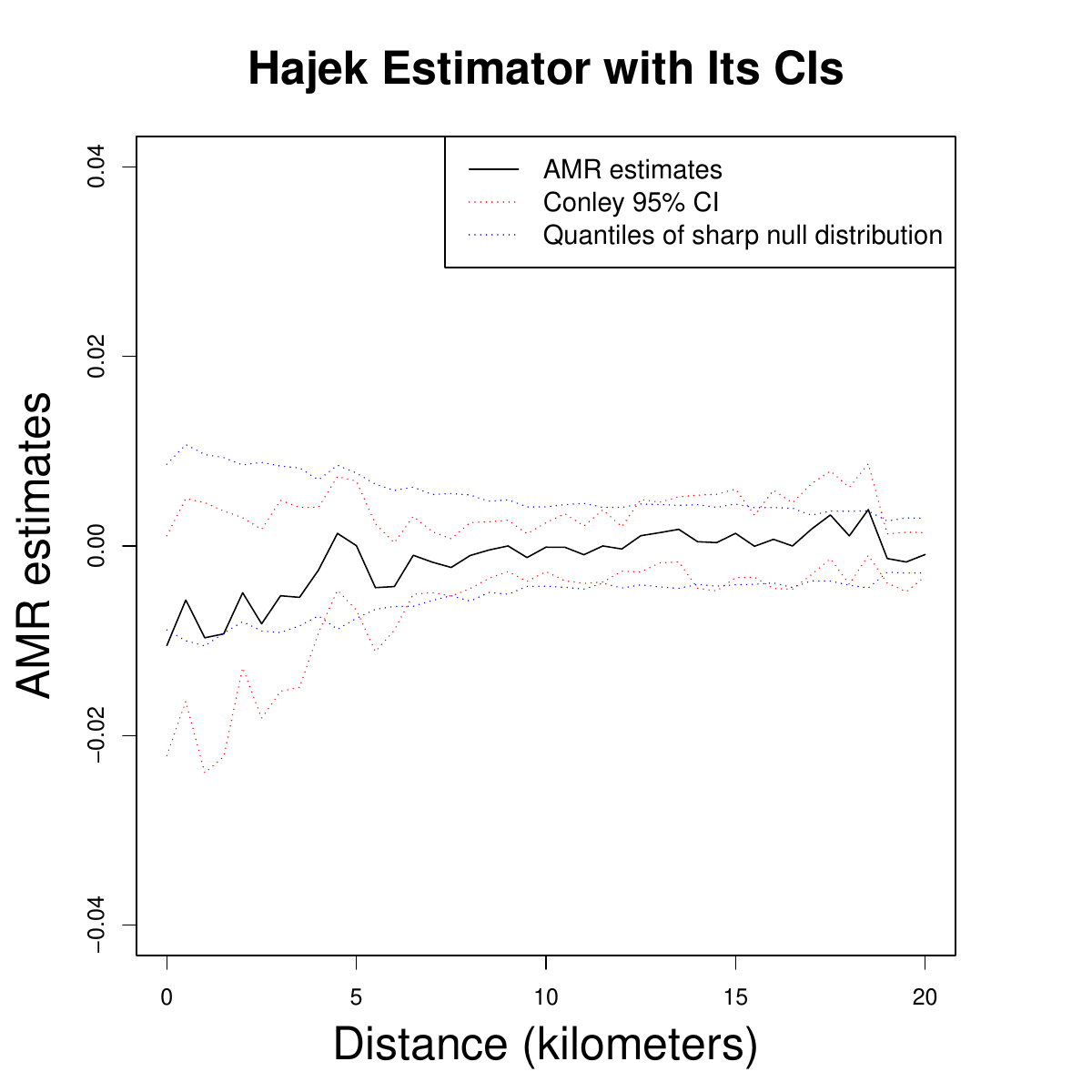}
\includegraphics[width=0.4\textwidth]{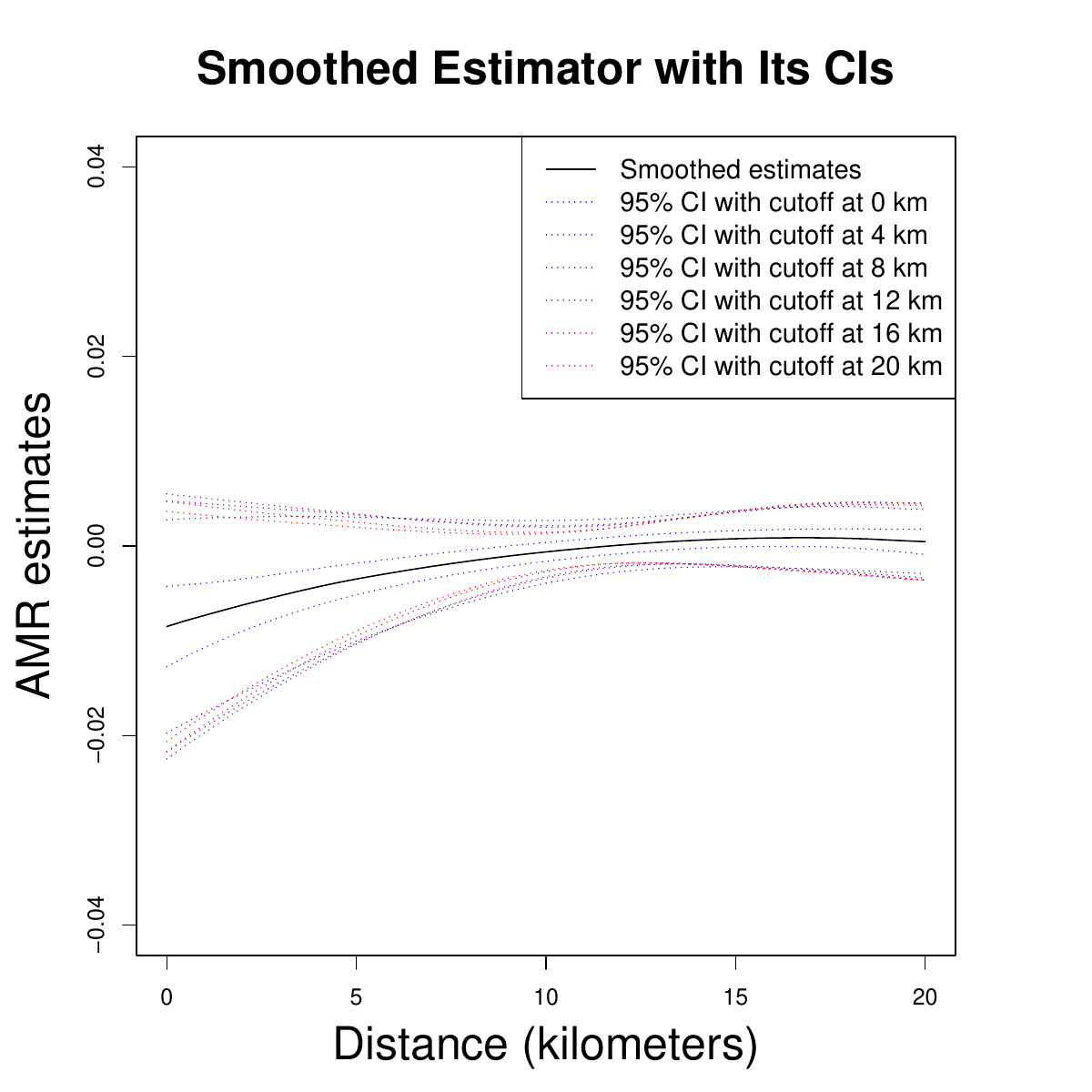}
\includegraphics[width=0.4\textwidth]{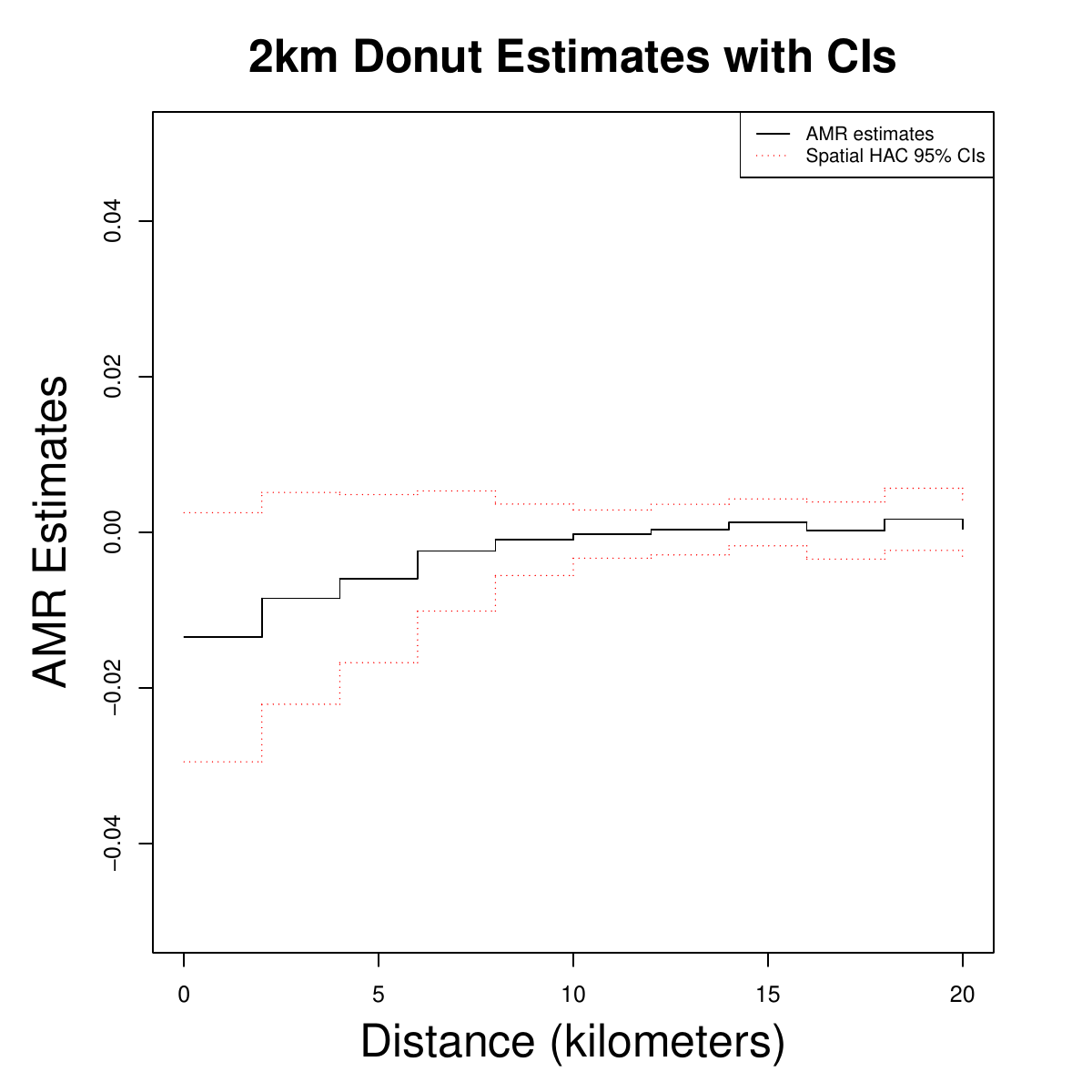}
\caption{The black solid line in the graph is the AMR estimate for each of the distance values $d$ from the Hajek estimator. The red dotted line is the 95\% confidence interval based on the spatial-HAC standard errors. The blue dotted line shows the 0.025 and .975 percentiles of the sharp null permutation distribution.}
\label{fig:exp-results}
\end{figure}

We display the estimates for distance values between 0 and 20 kilometers in Figure \ref{fig:exp-results}.  Similar to the analysis of the toy example, the black solid lines in Figure~\ref{fig:exp-results} are the AMR estimate for each distance value $d$ from the Hajek estimator. We estimate that the PES intervention generates effects that transmit outward from the treated villages. The effects gradually diminish in distance, and disappear when the distance to a treated village is larger than 10 kilometers. We find the same pattern in non-smoothed and smoothed estimates. We show how the confidence intervals for the smoothed estimates vary with the selected value of the cutoff. When cutoff equals 0, the confidence intervals ignore the uncertainties caused by interference. Hence, they are much narrower compared to the others. The bottom graph shows 2km ``donut'' estimates: we estimate the AMR using outcome raster outcomes within 2km bands, rather than values precisely at the edge of the circle. The results show a similar pattern as the two graphs above.

We test whether the cumulative effect between 1 and 2.5 kilometers are statistically significant using the {\tt SpatialEffectTest()} function:
\begin{center}
\begin{verbatim}
testResult <- SpatialEffectTest(AMRest, test.range = c(1/111, 2.5/111))
\end{verbatim}
\end{center}
For the cumulative effect between 1 km and 2.5 km, the 0.025 and .975 percentiles of the sharp null permutation distribution are $-0.0304$ and $0.0336$ (that is, a 3.04 percentage point reduction to a 3.36 percentage point increase in the cumulative probability of forest loss), respectively. The estimated cumulative effect between 1 km and 2.5 km is $-0.0321$ (i.e., a 3.21 percentage point reduction). The cumulative effect between 1 km and 2.5 km is thus significant with $p <0.05$. Beyond 2.5 km, spillover effects of the PES program appear to be negligible.

Thus, we find evidence for a spatial bandwagon effect that emanates up to 2.5 km outward beyond the intervention sites.  
These results are somewhat at odds with what \citet{Jayachandran267} conclude regarding spillover effects, although they did not attempt to estimate precisely the kind of spatial effect that is captured by the AMR.  The closest test that they conduct is to analyze whether there was less forest loss in control areas on the basis of the number of treated villages within 5 km.  
The key reason for why we find more compelling evidence is that our analysis is not restricted to forest loss within the village boundaries per se, which was that case for \citet{Jayachandran267}.  
Our analysis extends to forest areas that are outside the study villages too.  
An implication of our findings is that the between-village differences \citet{Jayachandran267} use to estimate the effect of the program may understate the overall impact of the PES intervention.  

\section{Conclusion}

This paper presents a robust, design-based approach to inference for effects in spatial experiments in environmental, health, agricultural, and other social applications.  Spatial experiments often exhibit complex displacement, bandwagon, and other spillover effects. 
These are forms of ``interference,'' which means that effects at a given point depend not only on the treatment of the nearest intervention site, but on the distribution of treatments over all intervention sites. Ignoring interference can result in mistaken conclusions. Mistakes can also result from relying on overly simplified models of spatial effects. The approach presented here makes careful use of the experimental design to make reliable inferences while remaining agnostic as to the precise nature of the outcome data generating process.  

The presentation here follows \citet{wang-etal-spatial} and focuses on randomized experiments.  Extensions to observational studies are possible in principle, so long as the treatment assignment mechanism abides by a process that is equivalent to what is described in \citet{wang-etal-spatial}.  We plan to extend the {\tt SpatialEffect} package to include covariate-conditional analyses that could broaden the range of possible observational applications.

We also offer detailed examples for applying the methods we present. These methods are encoded in the {\tt SpatialEffect} package for R, and we offer a step-by-step tutorial above on its use.  We then apply these tools to reanalyzing the spatial effects of an intervention that offered payments to promote forest conservation in Uganda. We uncover spatial effects that were not appreciated in the original analysis.  These effects suggest that the original analysis may have understated the overall impact of the program, and offer an example for why it can be crucial to account for interference in analyzing spatial experiments.

\end{document}